\documentclass[%
 reprint,
 amsmath,amssymb,
 prl,
 aps,
]{revtex4-2}

\usepackage{xr-hyper}
\usepackage{hyperref}
\usepackage{graphicx}% Include figure files
\usepackage{dcolumn}% Align table columns on decimal point
\usepackage{bm}% bold math
\usepackage[dvipsnames]{xcolor}
\usepackage{soul}
\usepackage{etoolbox}
\usepackage{soul}

\soulregister{\cite}{7} 
\soulregister{\ref}{7}  
\newtoggle{showEdits} 

\newcommand{\jtDelet}[1]{%
  \iftoggle{showEdits}{\textcolor{gray}{\st{#1}}}{}%
}

\begin{document}

\preprint{APS/123-QED}

\title{First Demonstration of Optical Feedback Control to Parametric Instability \\ at Advanced LIGO}% Force line breaks with \\

\author{Juntao Pan$^1$, Carl Blair$^1$, Vladimir Bossilkov$^2$, Jian Liu$^1$, Alex Adam$^1$, Chunnong Zhao$^1$, Li Ju$^1$, Anamaria Effler$^2$}
 
\affiliation{%
$^1$University of Western Australia, Crawley, Western Australia 6009, Australia
}%

\affiliation{%
$^2$LIGO Livingston Observatory, Livingston, LA 70754, USA
}%

\collaboration{LIGO Scientific Collaboration}%\noaffiliation
\date{\today}% It is always \today, today,

\begin{abstract}
Increasing the circulating power in gravitational-wave detectors to the megawatt level is essential for future sensitivity improvement, but this is critically limited by optomechanical parametric instabilities. Current mitigation strategies are projected to be inadequate against instabilities when circulating power reaches megawatt. Optical feedback offers a novel independent paradigm to mitigate parametric instability. In this Letter, we report the first demonstration of  optical feedback control in a full-scale gravitational wave detector.  We successfully suppress an unstable mode at 10.428 kHz, reducing the parametric gain from $R\approx 2$ to $R<0.02$. This work validates optical feedback control as an effective mitigation scheme for kilometer-scale interferometric gravitational-wave detectors, providing an effective strategy to allow detectors to reach the megawatt level.

\end{abstract}

\maketitle

%\tableofcontents

%\section{Introduction}

\textit{Introduction--}In the past decade, gravitational-wave astronomy has rapidly evolved into a rich observational science\cite{abac2025gwtc, abbott2017gw170817,palmese2024standard, abbott2020gw190412,abbott2020gw190521, abbott2020gw190814,abbott2021observation}. This evolution has been driven by continuous improvements in detector sensitivity. This enhancement enabled the observation of the event GW170817, which launched the era of multi-messenger astronomy and solved the origin of short gamma-ray bursts\cite{abbott2017gw170817,palmese2024standard}. The high signal-to-noise ratio signal GW250114 highlights this sensitivity enhancement. This signal amplitude was similar to the first detection GW1500914, but with \jtDelet{so much higher}a three time higher signal-to-noise ratio Hawking's area law could be verified to 4.4 $\sigma$.\cite{abac2025gw250114}.

The next frontier in gravitational-wave(GW) astronomy aims to illuminate previously invisible astrophysical phenomena, such as neutron star post-merger remnants\cite{maggiore2020scienceET,reitze2019cosmic}. Such detections require higher detector sensitivity, necessitating an increase in circulating laser power to the megawatt level. However, higher circulating power is challenging, introducing severe thermal and optomechanical problems. Optomechanical parametric instabilities(PI) is a runaway optomechanical oscillation driven by three-mode coupling among the fundamental optical field, a high-order transverse optical mode(HOM), and a mechanical mode of the test mass.

PI is a proven threat in advanced LIGO(aLIGO) at circulating powers of only $50$ kW\cite{evans2015observation}. The successful mitigation of PI has allowed aLIGO to reach a circulating power of $300\,$kW. This was achieved through a portfolio of techniques, including electrostatic damping (ESD)\cite{blair2017first}, acoustic mode dampers(AMD)\cite{biscans2019suppressing}, and thermal tuning of the three-mode detuning\cite{hardwick2020demonstration}. In particular, AMDs reduce the quality factors of mechanical modes over a broad frequency range ($\sim 12$–$80\,$kHz), thereby suppressing many potentially unstable modes and alleviating the control burden on ESD and thermal tuning. However, limitations of this cooperative strategy are already apparent, evidenced by the instabilities around $80.4\,$kHz that remain difficult to manage.  AMDs provide limited damping at $80\,$kHz, ESD has not been effective, probably due to poor spatial overlap between the ESD and the mechanical mode 
\jtDelet{These methods face effectiveness limitation at megawatt power levels. The high density of mechanical modes precludes thermal tuning, as detuning one instability inevitably results in others becoming unstable. Further, ESD becomes intractable because it requires a separate control loop for each of the tens to hundreds of unstable mechanical modes. Furthermore, AMD, if enhanced to provide sufficient damping, would introduce unacceptable thermal noise into the GW detection band. Therefore, a novel effective mitigation method is required.}

Looking ahead, next-generation detectors with over megawatt circulating power and $\geq100\,$kg test mass, are expected to experience more severe PI. Simulation predicts\cite{juntao_pan_ligo-t2600159-v1}, \jtDelet{after thermal tuning mitigation,} $\sim 8$ unstable modes in the $5$–$15\,$kHz \jtDelet{band needed to be suppressed}that AMD can not suppress without significant thermal noise penalty and will need to be suppressed by other means. \jtDelet{Enhancing AMD damping factor in this band would introduce unacceptable thermal noise into the GW detection band.} Moreover, the parametric gain required to be damped potentially reach $400$, exceeding the $\sim 100$ achievable from AMDs. These limitations motivates the development of new approaches that can operate alongside current methods.

Optical feedback(OFB) provides a robust, independent PI mitigation solution \jtDelet{that solves the problems of the current schemes }can be easily implemented in current and future detectors\cite{zhang2010enhancement,fan2010testing,bossilkov2024demonstration}.
\jtDelet{, with advantages in efficiency and simplicity}. This scheme works by injecting a control optical field(OFB HOM) that destructively interferes with the HOM that causes PI. OFB inherently requires less control loops than ESD, as a single OFB loop suppresses an optical mode that can be responsible for instability in many mechanical modes. The total number of optical modes that cause instability is \jtDelet{smaller by a factor of between 10 and 100 than }typically 10 to 100 times smaller than the total number of mechanical modes that become unstable. Tuning about 10 OFB control loops is manageable while tuning 100s of ESD control loops is not. \jtDelet{Furthermore, comparing to AMD, }OFB also introduces no additional thermal noise and negligible control noise. These characteristics make OFB an ideal candidate for PI suppression below 15 kHz \jtDelet{, serving as a vital complement.}in future detectors and to suppress instabilities like the $80\,$kHz mode in current detectors.

%\jtRevi{OFB inherently requires less control loops than ESD as a single OFB loop suppresses multiple unstable mechanical modes. Since a single high-order optical mode usually excites multiple unstable mechanical modes, suppressing that optical mode simultaneously stabilizes the entire unstable mechanical mode bunch. Moreover, unlike AMD, OFB introduces no mechanical dissipation and thus no thermal noise. Its residual control noise could be negligible once the instability is suppressed.}  

Historically, the complexity of spatially shaping the control field has hindered OFB implementation\cite{fan2010testing}. However, a recent experiment in an 80-m suspended cavity demonstrated that such shaping is unnecessary for low-order OFB HOMs\cite{bossilkov2024demonstration}. By exploiting intrinsic cavity coupling, the injected TEM$_{00}$ sideband converts into OFB HOM, effectively suppressing PI.

The feasibility of OFB in a full-scale, dual-recycling GW detector remained unproven. The key uncertainty is the magnitude of the coupling coefficient $G_0$, which quantifies the power conversion from the injected TEM$_{00}$ sideband into the OFB HOM. This reliance on mode conversion stands in direct conflict with detector designs optimized to minimize misalignment/mode-mismatch.

In this Letter, we report the first demonstration of OFB control of PI in a full-scale GW detector at the LIGO Livingston Observatory, successfully suppressing a 10.428 kHz instability. We subsequently evaluate the coupling coefficient $G_0$, providing information for OFB design of future megawatt detectors. We further discuss the feasibility of OFB to suppress $80\,$kHz instability in aLIGO, and OFB implementation to next-generation detectors.

%In this Letter, we report the first demonstration of OFB control of PI at LIGO Livingston Observatory. By successfully inducing and actively damping the instability with parametric gain of $R=1.9$ to $R<0.02$, we demonstrate the feasibility and effectiveness of OFB scheme. The damping ratio of two orders enables megawatt detector to experience equivalent PI severity of a $15\,$kW detector, promising to realize PI-free operation. This result \st{establishes} proves OFB to be an essential technology for PI mitigation in current and future GW observatories. 

%\section{Parametric Instabilities and Mitigation}

\textit{Parametric Instabilities--}The PI severity is described by the parametric gain\cite{evans2010general,evans2015observation}, $R$, given by
\begin{equation}
    R=\frac{8\pi Q_m P}{Mc\omega_m^2\lambda_0}\mathbf{Re}[G_n]B^2
    \label{eq:PImodel_Evan}
\end{equation}
Here, $Q_m$ is the quality factor of mechanical mode, $P$ circulating power, $M$ mass of test mass, $c$ speed of light, $\lambda_0$ laser wavelength, $\omega_0$ mechanical modes angular frequency, $G_n$ is the round-trip optical transfer function,$B^2$ spatial overlap between the three modes.  

When $R$ exceeds unity, the energy injection from the optical field overcomes mechanical dissipation, triggering PI and driving the exponential growth of the mechanical mode amplitude. The time constant of this growth is given as 
\begin{equation}
    \tau_{PI} = \frac{\tau_0}{1-R}
    \label{eq:tau_PI}
\end{equation}
where $\tau_0$ is the intrinsic ringdown time constant of the mechanical mode. If unmitigated, this instability saturates the control systems, halting detector operation.

\begin{figure}[!tb]
    \centering
    \includegraphics[width=0.85\linewidth,trim={0 0 0 0},clip]{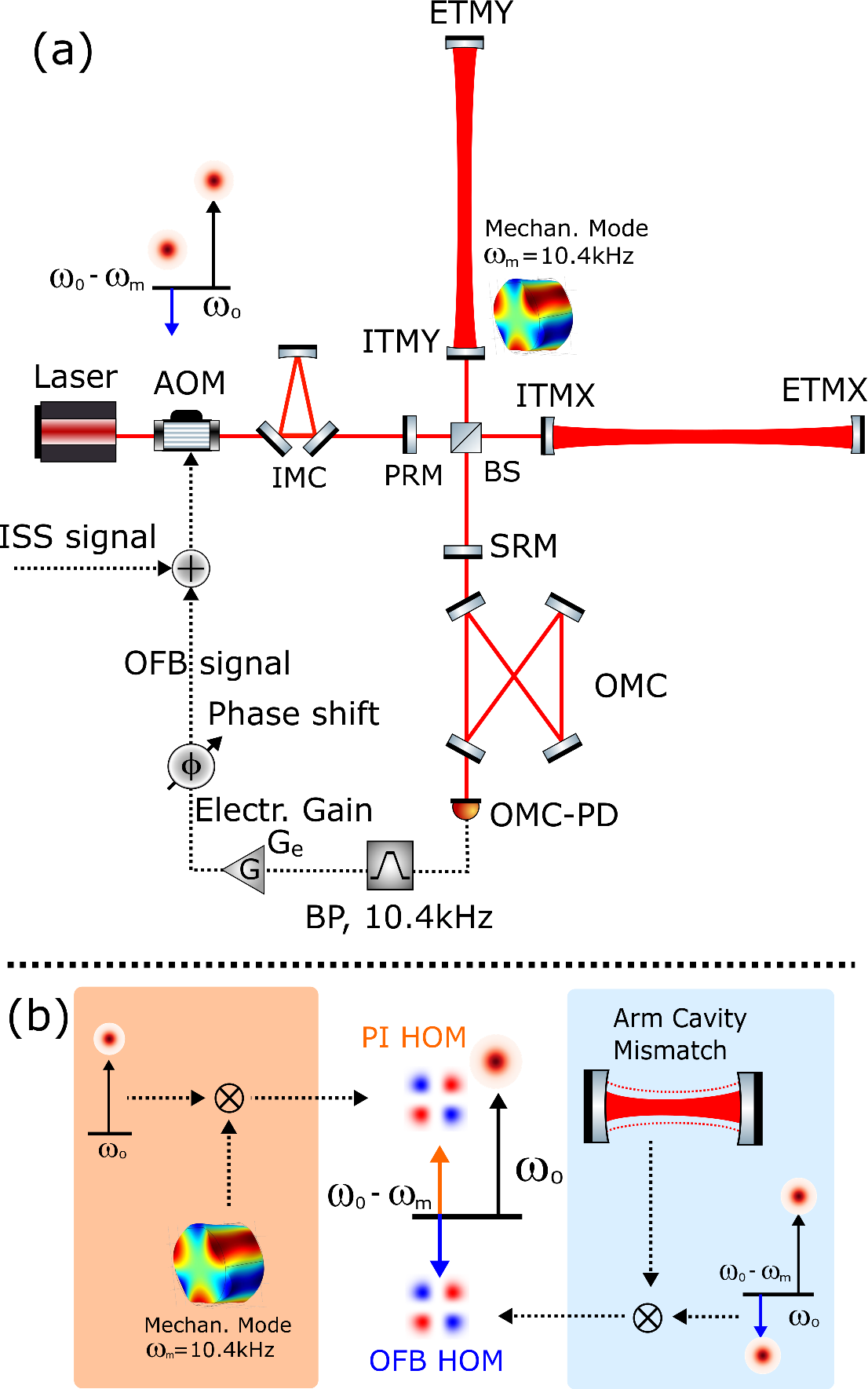}
    \caption{(a) Schematic of the optical feedback (OFB) control loop. The beat note between the fundamental TEM$_{00}$ mode and the PI-involved TEM$_{11}$ HOM is detected at the interferometer's main output port (OMC-PD). This error signal is processed and fed back to an acousto-optic modulator (AOM) to generate a TEM${00}$-profiled control sideband. (b) Principle of OFB scheme. The PI process involves the scattering of the fundamental mode into the PI HOM by the mechanical mode(orange shaded) . In OFB control, the injected control sideband is converted into a TEM$_{11}$ OFB HOM via an mode-mismatch(blue shaded region); OFB HOM then destructively interferes with the PI HOM to suppress the instability. }
    \label{fig:setup}
\end{figure}

%\section{Experiment Setup}
\textit{Experiment Setup--}To induce PI, we increased the ETMY and ETMX ring heater powers by $+0.8\,$W /$+0.5\,$W, respectively.  The ring heater is a circular heating element that heats a ring around the barrel of the mirror. The ring heater induces a thermal gradient that reduces the mirror radius of curvature. The change in radius of curvature of the mirrors tunes the transverse optical mode frequency spacing to overlap with the mechanical mode frequency at $10.428\,$kHz. This changes the round trip optical transfer function $G_n$. The parametric gain is maximized when the transverse mode frequency beat with the carrier light is equal to the mechanical mode frequency \cite{evans2010general}.

\jtDelet{The $10.428\,$kHz mechanical mode scatters the fundamental mode into the TEM$_{11}$ high-order mode(PI HOM).}The $10.428\,$kHz mechanical mode scatters a fraction of the fundamental mode power into the $\text{TEM}_{11}$ mode(PI HOM), which resonates within the interferometer and further driving the mechanical mode to cause PI. The OFB control targets this TEM$_{11}$ mode for suppression, in the scheme shown in FIG.~\ref{fig:setup}. The optical beatnote between fundamental mode and PI HOM serves as error signal and is detected at the interferometer's main output port (OMC-PD). The output mode cleaner possesses a linewidth exceeding $600\,$kHz, and the $10.4\,$kHz signal remains readily observable. However, the exact coupling mechanism is not well understood. The error signal voltage, $V_{\text{err-sig}}$, is proportional to the mechanical mode's amplitude, \jtDelet{$A_m$}$A$,

\begin{equation}
   % \jtDelet{ V_\text{err-sig}\propto P_{\text{bt}} = E_{00}^2\frac{\pi A_m}{\lambda_0} \cos(\omega_mt+\phi_m) }
   V_\text{err-sig} \propto E_{00}^2 A \cos(\omega_mt+\phi_m)
    \label{eq:BeatPowPropAm}
\end{equation}
where $\omega_m = 2\pi \times 10.428\,$kHz is the mechanical angular frequency, $E_{00}^2$ is the circulating power of the fundamental mode, and $\phi_m$ denotes the relative phase between the fundamental mode and the PI HOM.

Subsequently, the error signal is electronically processed and used to drive an acousto-optic modulator to generate a TEM$_{00}$-profiled control sideband. This sideband is converted into the TEM$_{11}$ (OFB HOM) in the arm cavity through mode mismatch. The error signal is phase shifted to realize destructive interference with the PI HOM.

%\section{Results}

\begin{figure}[!tb]
    \centering
    \includegraphics[width=0.95\linewidth]{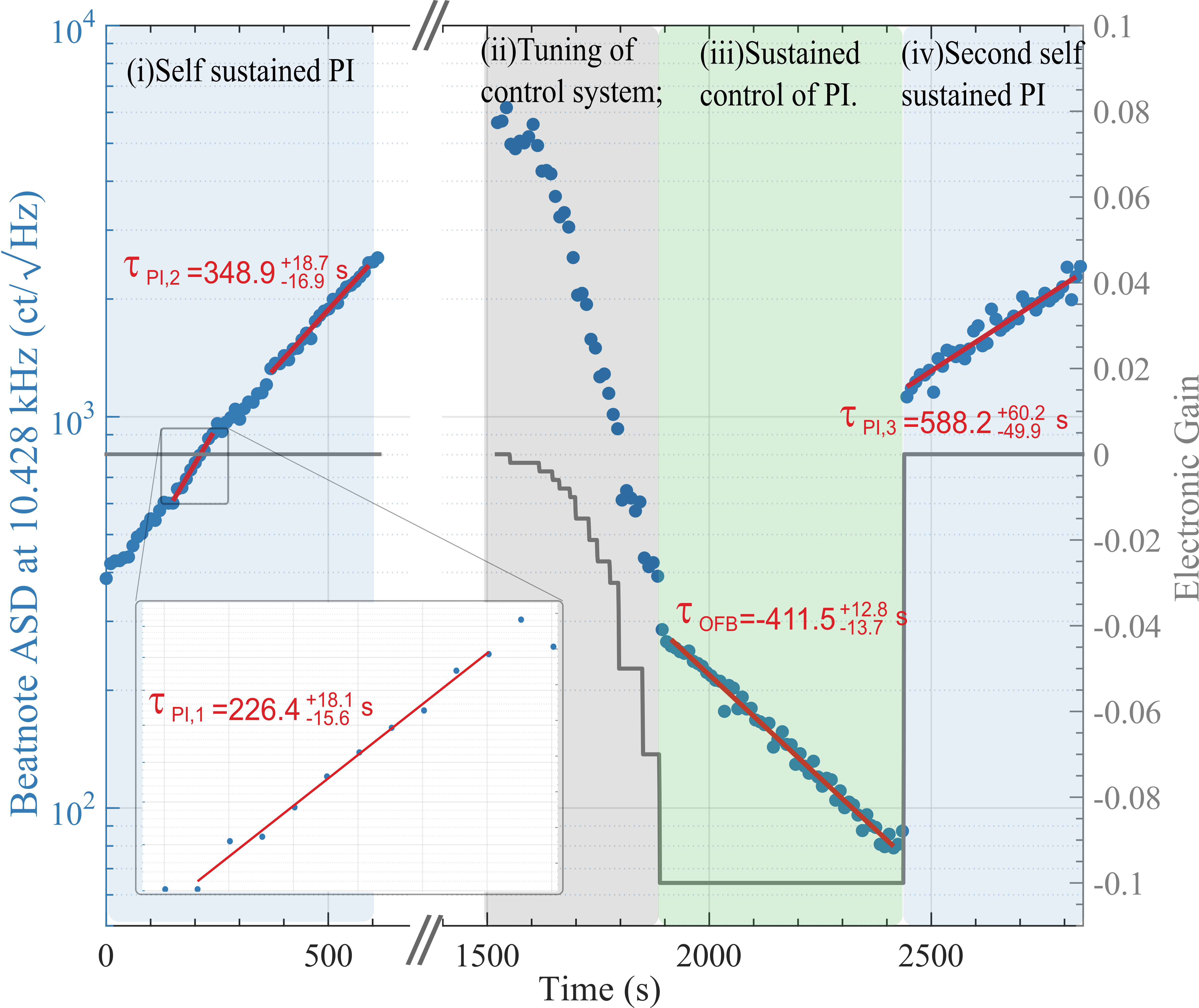}
    \caption{Demonstration of OFB control to PI. The error signal amplitude spectral density (ASD) is tracked (blue dots, left axis), which is proportional to the mechanical mode amplitude. The experiment proceeds in four phases:(i)self-sustained PI; (ii) tuning of the OFB control system; (iii) sustained control of PI, and (iv) immediate regrowth upon OFB disengagement. The electronic gain (gray line, right axis) is engaged at $t=1500\,$s and increased gradually; the phase is tuned to ensure the control sideband is out of phase to PI HOM. Once the electronic gain reaches $-0.1$, the OFB loop operates\jtDelet{autonomously} without manual intervention. Red lines indicate exponential fits to the mechanical amplitude growth/decay within the intervals labeled PI-1, PI-2, PI-3, and OFB. \jtDelet{This immediate re-emergence at $2400\,$s provides definitive proof of the OFB control's efficacy in successfully suppressing the PI.}}
    \label{fig:MainRes}
\end{figure}

\textit{Results--}FIG~\ref{fig:MainRes} shows the amplitude of the beatnote signal as a function of time. This figure shows four distinct phases, (i)Self sustained PI; (ii)Tuning of the control system; (iii)Sustained control of PI. (iv)Followed by another Self sustained PI period.

For $t < 1500\,$s, the instability was induced and left unmitigated, growing by an order of magnitude. At $1500$ s, we engaged the OFB control. The electronic gain $G_e$ was gradually increased to prevent actuator saturation. Upon reaching $G_e=-0.1$, the OFB HOM strongly suppress the instability, converting the exponential growth into exponential decay from $1900$ to $2400\,$s. To verify this active control, we intentionally disengaged OFB at $2400\,$s. The sudden cessation of interference resulted in the sharp rise in signal (70 to 1000 $\text{ct}/\sqrt{\text{Hz}}$) and the resumed instability. 

To quantify OFB effectiveness, we analyze the parametric gain $R$. As shown by Eq.(\ref{eq:tau_PI}), more severe instabilities have a higher $R$ and a shorter growth time constant, $\tau_\text{PI}$.  We selected four time intervals and fitted their time constant ($\tau_\text{PI/OFB}$), as shown in FIG.\ref{fig:MainRes}.  The intrinsic mechanical ring-down time, $\tau_0 = -406.3 \pm 10.1$ s, was measured in a separate experiment with much reduced optical power and high detuning to minimize the optomechanical interaction. The parametric gains of each intervals are then calculated and summarized in TABLE.~\ref{tab:FreqShiftTrack}.  

The observed gains varied from $R_\text{PI,1}=2.79$ to $R_\text{PI,3}=1.69$. The variation attributed to the thermal drift of the TEM$_{11}$ resonant frequency. The parametric gain at the moment of OFB engagement is therefore somewhat uncertain, but must falls between the values of $R_\text{PI,2}=2.16$ and $R_\text{PI,3}=1.69$ measured before and after the control. The thermal drift in the HOM resonant frequency is modeled to provide a better estimate of the parametric gain at the time of control.

\begin{table}[!b]%The best place to locate the table environment is directly after its first reference in text
\renewcommand{\arraystretch}{1.3}
\caption{\label{tab:FreqShiftTrack}
The observed parametric gains $R$ and thermal frequency shifts for the four selected time intervals. The frequency shifts are relative to the PI,1 stage.}
\begin{ruledtabular}
\begin{tabular}{cccc}
\textrm{Time Interval}  & \textrm{Notation}& 
\textrm{ Param. Gain$R$}&
\textrm{ Freq. Shift(Hz) }\\
\colrule
$150-250\,$s  &       PI, 1   &       $2.79^{+0.18}_{-0.17}$  &       $0\pm0.56$                    \\
$400-600\,$s  &            PI, 2   &       $2.16^{+0.09}_{-0.09}$  &       $2.35\pm0.82$                 \\
$1910-2420\,$s   &              OFB   &        $0.013^{+0.008}_{-0.006}$  &         $4.19 \pm{0.59}$                          \\
$2440-2830\,$s  &             PI, 3   &       $1.69^{+0.08}_{-0.08}$  &       $6.26\pm0.63$                
\end{tabular}
\end{ruledtabular}
\end{table}

%\section{R Variation Modeling}

\begin{figure}[!tb]
    \centering
    \includegraphics[width=0.9\linewidth]{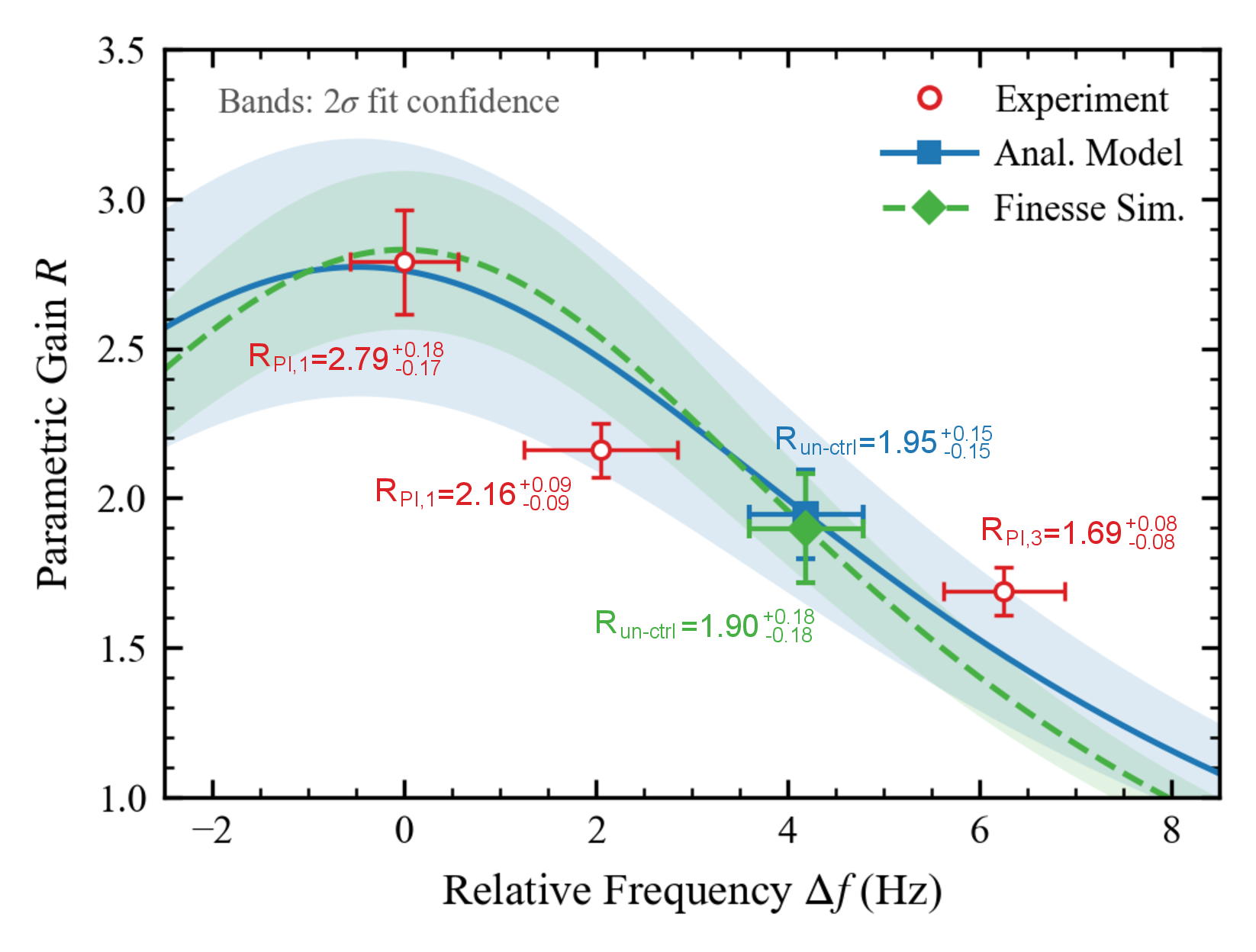}
    \caption{Parametric gain $R$ as a function of the TEM$_{11}$ mode thermal frequency drift $\Delta f$. Experimental data (red dots) are compared with the analytical (blue curve) and Finesse (green curve) models. The models indicate that at the moment of OFB engagement, the control system confronted an unsuppressed parametric gain of $R=1.90$ (Finesse) or $R=1.95$ (analytical). The uncertainty of the unsuppressed gain only counts the uncertainty of frequency shift. Shaded regions denote confidence intervals derived from a Monte-Carlo analysis ($10^{4}$ iterations), sampling the uncertainties in both $R$ and the frequency shift $\Delta f$ and re-fitting both models respectively. }
    \label{fig:model_RVar}
\end{figure}

\textit{Model Variation in Parametric Gain--} The thermal non-equilibrium caused the ongoing frequency drift and variation in parametric gain.  We tracked this frequency drift of TEM$_{11}$ by fitting the power spectral density profile of the OMC transmission signal that shows the TEM$_{11}$ field driven by noise. \jtDelet{illustrated in FIG.\ref{fig:HOMDrift}.}The estimated average frequency shifts are summarized in Table.\ref{tab:FreqShiftTrack} , while the tracking method is detailed in the Supplementary Sec.\ref{supp-supsec:ThermDriftTrack}.

The optical response, $\textbf{Re}[G_n]$ in Eq.\eqref{eq:PImodel_Evan}, characterizes the $R$ variation induced by this frequency drift. As detailed in the Supplementary Sec.\ref{supp-supsec:GnProf_fit}, we exploited both an analytical model and a numerical Finesse-3 model, to evaluate $\textbf{Re}[G_n]$ profile and parametric gain. Subsequently these two models are fitted with the experimental data to evaluated the unsuppressed parametric gain when OFB engaged. The models predict parametric gains of 1.95(Analytic) and 1.90(Finesse) at the time of control as shown in FIG.~\ref{fig:model_RVar}.

\jtDelet{
Each PI stage was segmented into several 20-second intervals. The PSD of each interval was fitted with a Lorentzian function to extract the peak frequency. The average frequency and standard deviation for each PI stage were calculated from these fits. This procedure was repeated for all four time intervals. The average frequency shifts summarized in Table \ref{tab:FreqShiftTrack}.
}

\jtDelet{
The frequency drift changes the parametric gain via a change in optical gain( $\textbf{Re}[G_n]$ in Eq.\ref{eq:PImodel_Evan}). In Supplementary Material, we exploited both an analytical model and a numerical Finesse-3 model, to evaluate $\textbf{Re}[G_n]$ profile and parametric gain. Subsequently these two models are fitted with the experimental data to evaluated the unsuppressed parametric gain when OFB engaged. The models predict parametric gains of 1.95(Analytic) and 1.90(Finesse) at the time of control as shown in FIG.~\ref{fig:model_RVar} . 
}

%\section{Mismatching Coefficient}

\textit{Mismatching Coefficient--}The OFB scheme relies on a fortuitous coupling between the input fundamental mode and the high order mode in the arm cavity. A larger coupling gives a larger dynamic range actuator. This coupling is calculated to be $G_0=$\jtDelet{$3.11\times10^{-3}\,$}$1.34\times 10^{-4}\,$W/W in the Supplementary Sec.\ref{supp-supsec:MismatchG0}.

The OFB HOM is amplified within the interferometer with a gain of $\textbf{Re}[G_n]=863.58$ informed by the analytic model, $844.93$ from Finesse-3 model. This high gain compensates the low coupling efficiency, and provides adequate dynamic range to suppress PI that has grown by 10 above the quiescent level in this experiment.

\begin{figure}
    \centering
    \includegraphics[width=0.88\linewidth]{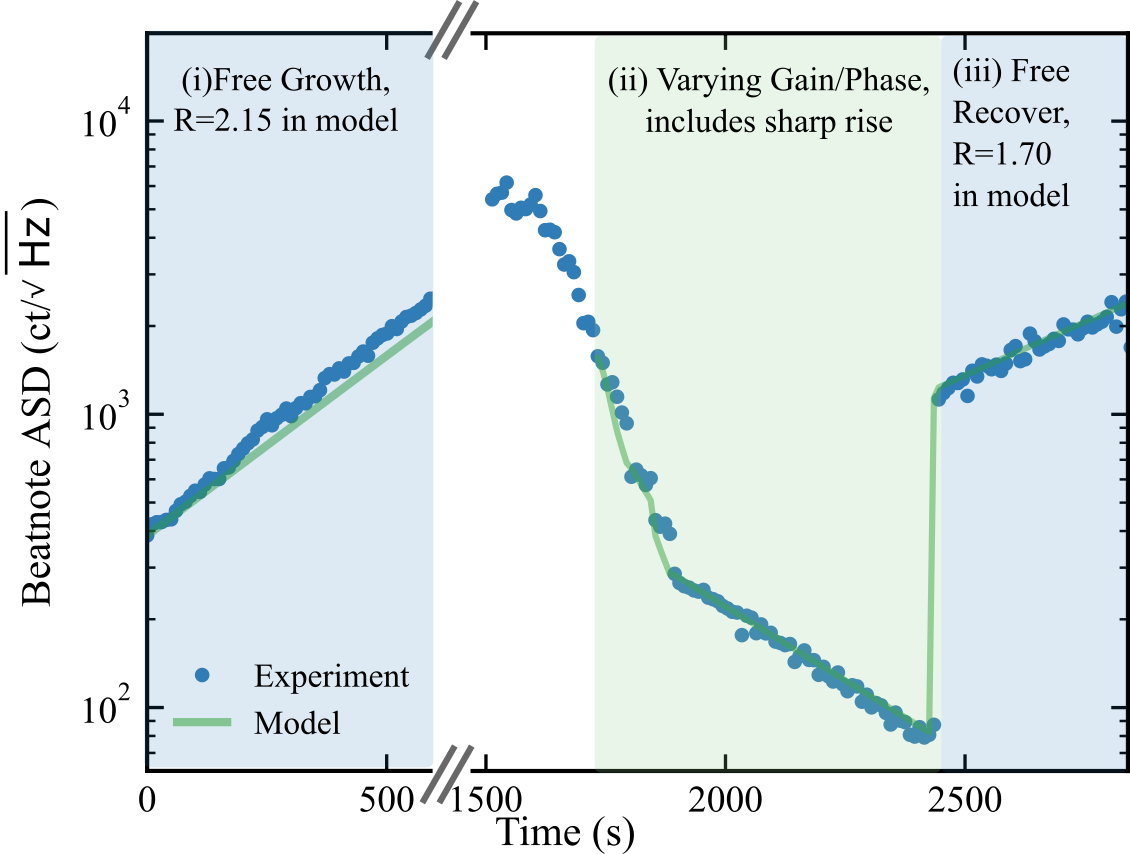}
    \caption{Time evolution comparison between experimental data(blue dots) and theoretical model(green curve). In the two free-evolution stages (blue-shaded regimes), the model accurately predicts the parametric growth using fixed gain of $2.15$ and $1.70$, respectively. The OFB-engaged simulation begins at $1733\,$s, where the electronic phase is experimentally determined and tuned within a small range. The gain and phase of control became fixed values when sustainable control achieved. The whole theoretical trace is generated utilizing the derived nonlinear framework, which notably replicates the sharp rise observed at approximately $2400\,$s. }
    \label{fig:Model_TimeEvolu}
\end{figure}

\textit{Time Evolution--} We develop the theory framework to model the dynamics of mechanical mode amplitude under OFB engagement in dual-recycling interferometer,
\begin{equation}
\begin{aligned}
  \dfrac{dA}{dt} &= \frac{\gamma_m}{2} \left( R_0 |A_c|^2 \mathcal{L}_{\mathrm{opt}} - 1 \right) A - k_{NL} A^3  \\
  &k_{NL} =\gamma_m R_0 |A_c|^2 \frac{\kappa_S}{\kappa_0} \mathcal{L}_{\mathrm{opt}}^2
  \label{eq:Resu_MechAmpEvo} 
\end{aligned}
\end{equation}
whose detailed derivation is included in Supplementary Sec.\ref{supp-supsec:TheoDeri_TimEvo}. The validity of our theoretical framework is confirmed by comparing the numerically integrated trace and the experimental data, shown in FIG.\ref{fig:Model_TimeEvolu}.

%\section{Conclusion}
\textit{Conclusion--} We demonstrated the suppression of parametric instability with $R\approx2$ using a OFB control scheme, which successfully reduced the net PI gain to $R<0.02$ in the aLIGO detector. This two-order-of-magnitude suppression capability implies that the PI in future detectors with 1 MW circulating power could be controlled to a level equivalent to that of a detector with a $\sim 10\,$kW circulating power, effectively rendering the current detectors PI free at the megawatt power level. This study provides a foundational demonstration of optical feedback control that can be used to design and implement this novel control strategy more widely. This result establishes OFB as a powerful tool, paving the way for next-generation megawatt-scale GW observatories that will illuminate previously invisible astrophysical phenomena.

\textit{Discussion--} \jtDelet{The control bandwidth in this demonstration was limited by a $4\,$ms latency from computer communication. Consequently, we restricted the OFB implementation to the suppression of a single mechanical mode.Reducing this latency will extend the control bandwidth, enabling the simultaneous suppression of multiple unstable mechanical modes.} The control bandwidth in this demonstration was restricted to a single mechanical mode due to a $4\,$ms inter-computer latency within the temporary experimental setup. However, the nominal inter-computer delay within the standard aLIGO control system is  $\sim 61\,\mu$s. \jtDelet{Transitioning to this native system} A properly designed system will eliminate current bandwidth restrictions, enabling the simultaneous suppression of multiple unstable mechanical modes.

The OFB effectiveness is strongly tied to the resonant amplification of the HOM within the interferometer, $G_n$. This gain is sensitive to both signal recycling cavity (SRC) configuration and the thermal state of the arm cavities.

We here discuss the applicability of OFB to suppressing the $\sim 80\,$kHz instability in aLIGO. Notably, this instability is predominantly driven by interactions with 1st-order HOMs of the second free spectral range, a condition that favors the implementation of OFB. The limitation arises from the attenuation of the injected sideband by the input mode cleaner and the power recycling cavity. For an $80\,$kHz sideband, the total attenuation is $-39\,$dB, compared to $-7.2\,$dB for the $10.4\,$kHz sideband used in this work.\jtDelet{ Despite this high attenuation,} Achieving a sideband power,\jtDelet{  injected into arm cavity,} comparable to the present experiment requires\jtDelet{ only} an AOM driving voltage of $\sim 1\,$mV. This remains well within practical limits, and still much less than the signal from intensity stabilization system(ISS),  $V_\text{ISS}\sim0.5\,$V. 

The setup here confirms OFB effectiveness. Future detector may require a larger dynamic range, i.e. more power injected in to the HOM of the arm cavity.\jtDelet{however may require adaptations to ensure sufficient dynamic range for further detector.} The primary limitation of the existing setup arises from generating the control field via ISS AOM and injecting it through interferometer's main input. To prevent disruption of the ISS loop,  the OFB modulation depth must remain small, $V_{\text{OFB}} \ll V_{\text{ISS}}$, which caps the maximum power of the control sideband. Furthermore, before reaching the target arm cavity, the control sideband undergoes attenuation as it passes through the input mode cleaner(IMC) and power recycling cavity(PRC). Together with the low scatter coefficient from sideband to OFB HOM, these factors restrict the dynamic range of OFB. These constraints can be bypassed by sensing the error signal and injecting the control field through the back surface of the end test masses. This alternative injection scheme circumvents AOM power limitations and avoids the IMC/PRC attenuation. Sensing the error signal at the ETM transmission introduces a signal-to-noise ratio challenge. While the intracavity circulating power is at megawatt levels, the power of the beatnote is over ten orders of magnitude weaker. Detecting this faint signal requires filtering out the dominant carrier field, and this can be achieved by implementing a Mach-Zehnder interferometer.

\textit{Acknowledgments--} The authors acknowledge the entire LIGO Scientific Collaboration for their wide ranging expertise and contributions. The authors acknowledge LIGO Science Collaboration colleagues for their contributions to the detectors. This material is based upon work supported by NSF's LIGO Laboratory which is a major facility fully funded by the National Science Foundation. LIGO was constructed by the California Institute of Technology and Massachusetts Institute of Technology with funding from the National Science Foundation, and operates under cooperative agreement PHY-0757058. aLIGO was built under award PHY-0823459. J.Pan is supported by the Australian Research Council under the ARC Centre of Excellence for Gravitational Wave Discovery, Grant No.CE230100016.  J. Pan acknowledges the LSC Fellows program for supporting the research at the LIGO Livingston Observatory.

%\newpage

\bibliography{ref}% Produces the bibliography via BibTeX.

\end{document}